\documentclass{iau}

\usepackage{amsmath}
\usepackage{graphicx}
\usepackage{placeins}

\usepackage[dvipsnames]{xcolor}

\begin{document}

\lefttitle{Peralta \& Vieytes}
\righttitle{Unsupervised ML for Spectral Line Selection}

\jnlPage{1}{4}
\jnlDoiYr{2026}
\doival{10.1017/xxxxx}

\journaltitle{Unraveling the Joint Lives of Stars and Exoplanets}
\aopheadtitle{Proceedings IAU Symposium}
\volno{408}
\editors{G. Buldgen, A. Vidotto \& A. Miglio, eds.}


\title{Chromospheric sensitivity of stellar spectral lines: an unsupervised machine learning classification}

\author{Juan I. Peralta$^{1,2}$ and Mariela C. Vieytes$^{1,2,3}$}
\affiliation{$^1$Instituto de Astronom\'ia y F\'isica del Espacio (IAFE, CONICET--UBA); $^2$Laboratorio de F\'isica, Universidad Nacional de Tres de Febrero (UNTREF); $^3$Ciclo B\'asico Com\'un, Universidad de Buenos Aires (UBA); Buenos Aires, Argentina\\
\email{mariela@iafe.uba.ar; jperalta@iafe.uba.ar}}

\begin{abstract}
Stellar magnetic activity alters thousands of spectral lines, limiting high-precision radial-velocity measurements, abundance analyses, and planetary characterization. We present a preliminary classification of 6403 atomic lines in synthetic visible spectra from a sequence of NLTE semi-empirical dG2 atmospheric models with increasing chromospheric heating. Principal-component analysis shows that the dominant component traces overall response amplitude, while the second distinguishes early from late responders. DBSCAN applied in the full nine-dimensional response space identifies a dense stable core and an activity-sensitive non-core group comprising about 11\% of the lines. The stable core provides candidate lines for activity-insensitive measurements, while sensitive lines provide candidate activity diagnostics. Sensitive transitions tend to have low lower-level energies, although the populations overlap, making lower-level energy a statistical discriminator rather than a line-by-line predictor. A spectral sensitivity map shows that small average variations can hide nonlinear or compensating responses, demonstrating the value of the complete line-response trajectory for classification.
\end{abstract}


\maketitle

\section{Models and line-response metrics}

Chromospheric activity induces measurable variations in thousands of spectral lines in solar-type stars, and those variations limit how well stellar and planetary signals can be separated. We therefore move from line-by-line examples to a global classification of activity-induced line-response patterns, identifying both activity-stable lines and candidate chromospheric diagnostics.

Using the sequence of NLTE semi-empirical G2-dwarf atmospheric models computed by \cite{Vieytes2025} (hereafter VZB25), we classify 6403 atomic lines according to their response patterns across the chromospheric-heating sequence. VZB25 showed that chromospheric heating can introduce secondary chromospheric contributions to line formation. Here we extend that analysis by asking which lines behave similarly across the full activity sequence, which remain stable, which become activity-sensitive, which physical properties distinguish the resulting groups, and where those properties stop being sufficient.

Starting from the quiet solar model of \cite{Fontenla2015}, recomputed with an updated Fe~I atomic model, VZB25 built a sequence of models by increasing the temperature of their chromospheres. The photospheric structure is the same in all of them, except for the temperature minimum, which rises and shifts inward; the plateau rises from 6140 to 7340~K across the sequence. This increase represents the mechanical energy deposition that strengthens the emission of the Ca~II K line core. The visible spectra were computed with the SSRPM code suite \cite{Fontenla2016}, described in VZB25, which treats 52 neutral and low-ionization species in non-LTE.

The present sample contains 6403 atomic transitions between 3300 and 7020~\AA, spanning 22 elements and 52 ions. It includes the 13 H\,I Balmer lines of this range, computed by SSRPM and present in the synthetic spectra but not listed individually in the line table of VZB25. 

For each transition $i$ and each active model $j$, the line-core response relative to the quiet model is quantified by RD, as defined by VZB25. To remove its mathematical asymmetry, we use LRD as the input representation for PCA and clustering:
\begin{equation}
 \mathrm{RD}_{i,j}=100\,\frac{I_{i,j}-I_{i,1701}}{I_{i,1701}},
 \qquad
 \mathrm{LRD}_{i,j}=\log_{10}\left(1+\frac{\mathrm{RD}_{i,j}}{100}\right).
 \label{eq:rdlrd}
\end{equation}

\begin{figure}
\includegraphics[width=0.49\textwidth]{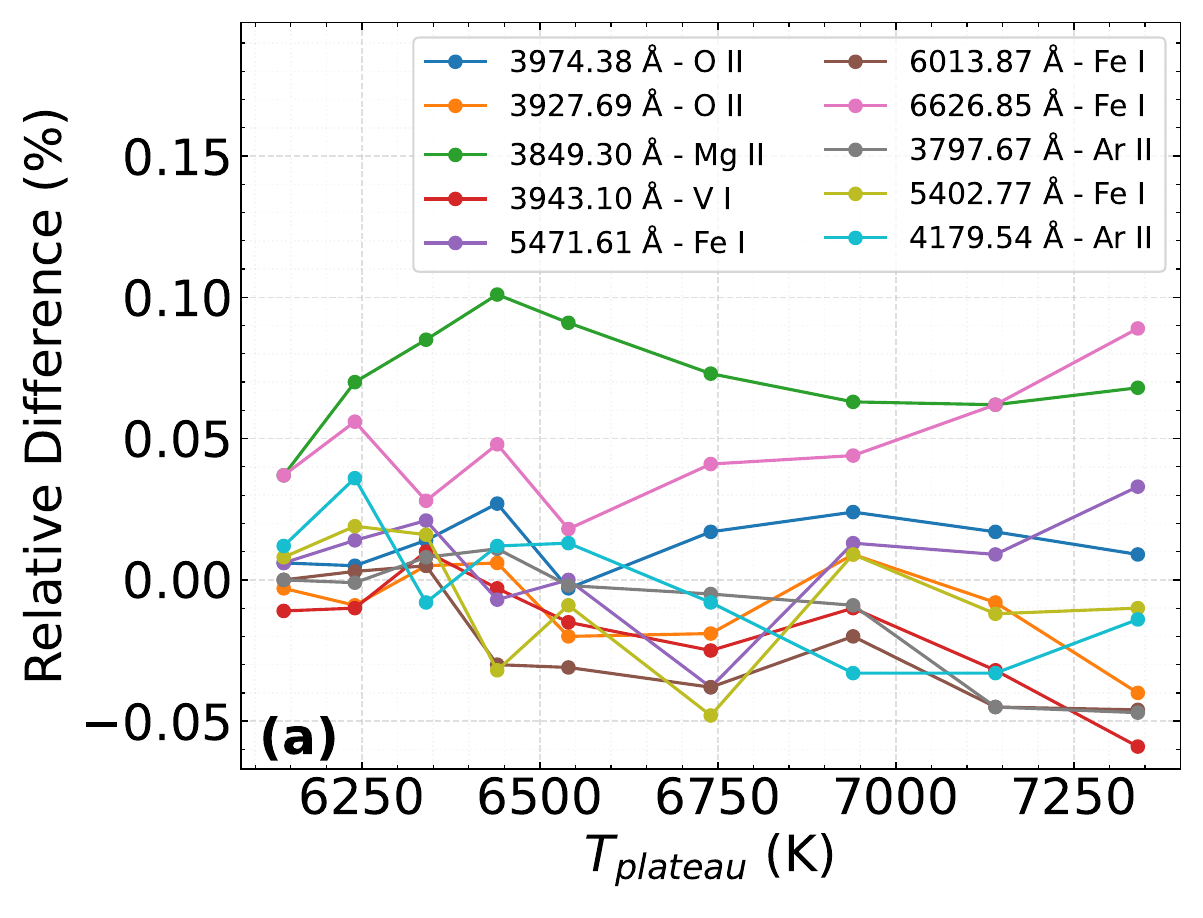}\hfill
\includegraphics[width=0.49\textwidth]{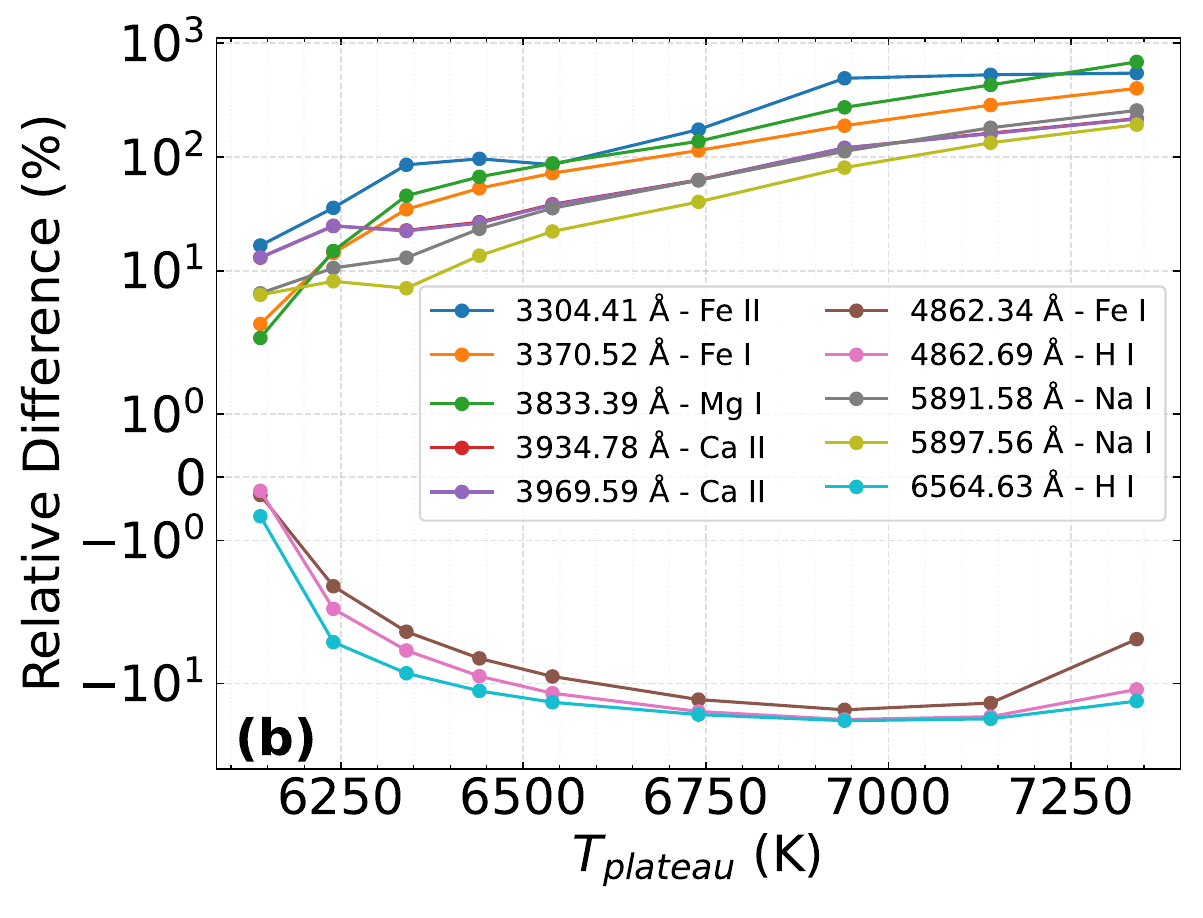}
\vspace{-0.5em}
\caption{Line trajectories along the model sequence. \textbf{(a)} Representative DBSCAN stable-core lines. \textbf{(b)} Representative DBSCAN-sensitive lines with positive and negative responses.}
\label{fig:models}
\end{figure}

Each line is represented by its full nine-model LRD trajectory, which preserves the amplitude and the shape of its response across successive increments of chromospheric heating instead of collapsing the response to a single value. Representative trajectories are shown in Figure~\ref{fig:models}. Stable lines remain nearly flat across the sequence, while sensitive lines can show positive or negative responses and can develop gradual, threshold-like, or nonlinear behaviour.

\section{Classification of line-response trajectories}

\begin{figure}
\centering
\includegraphics[width=0.49\textwidth]{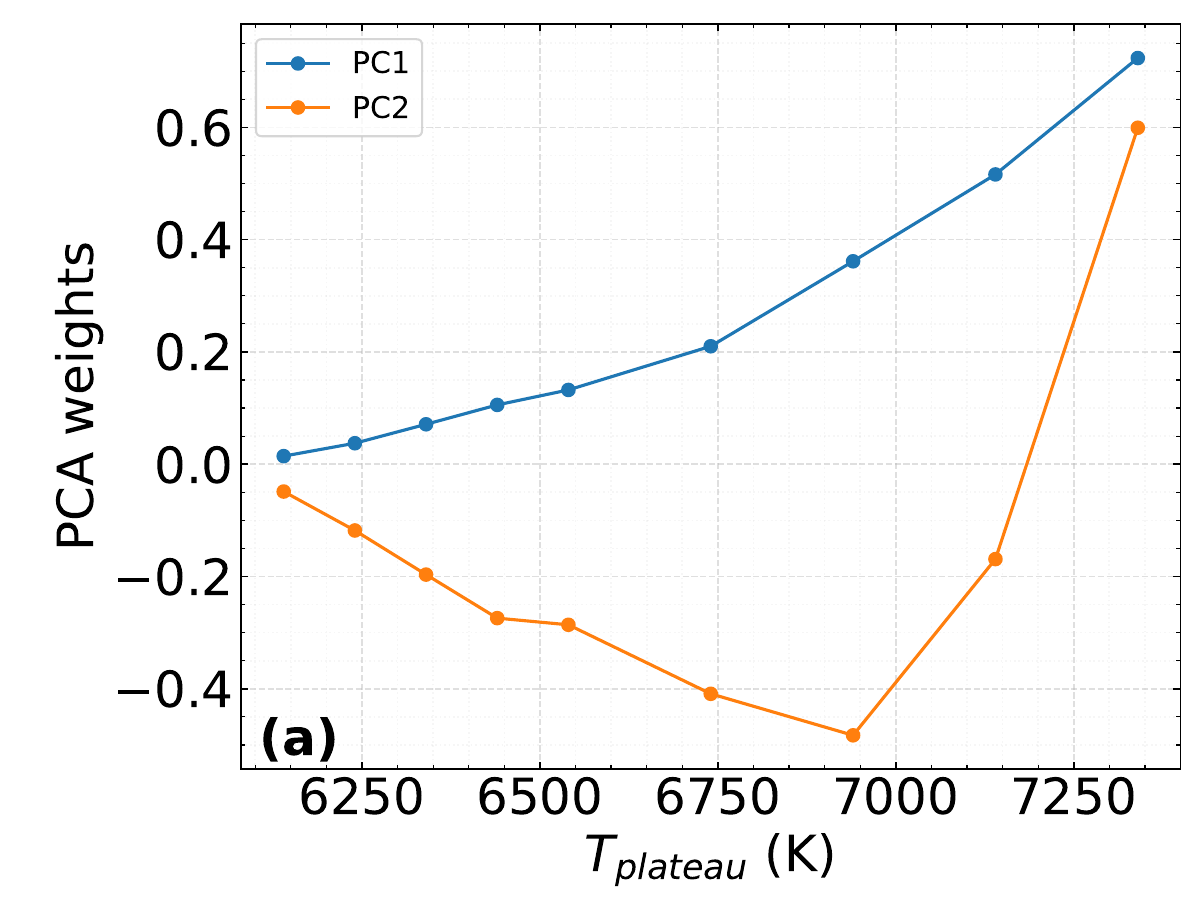}\hfill
\includegraphics[width=0.49\textwidth]{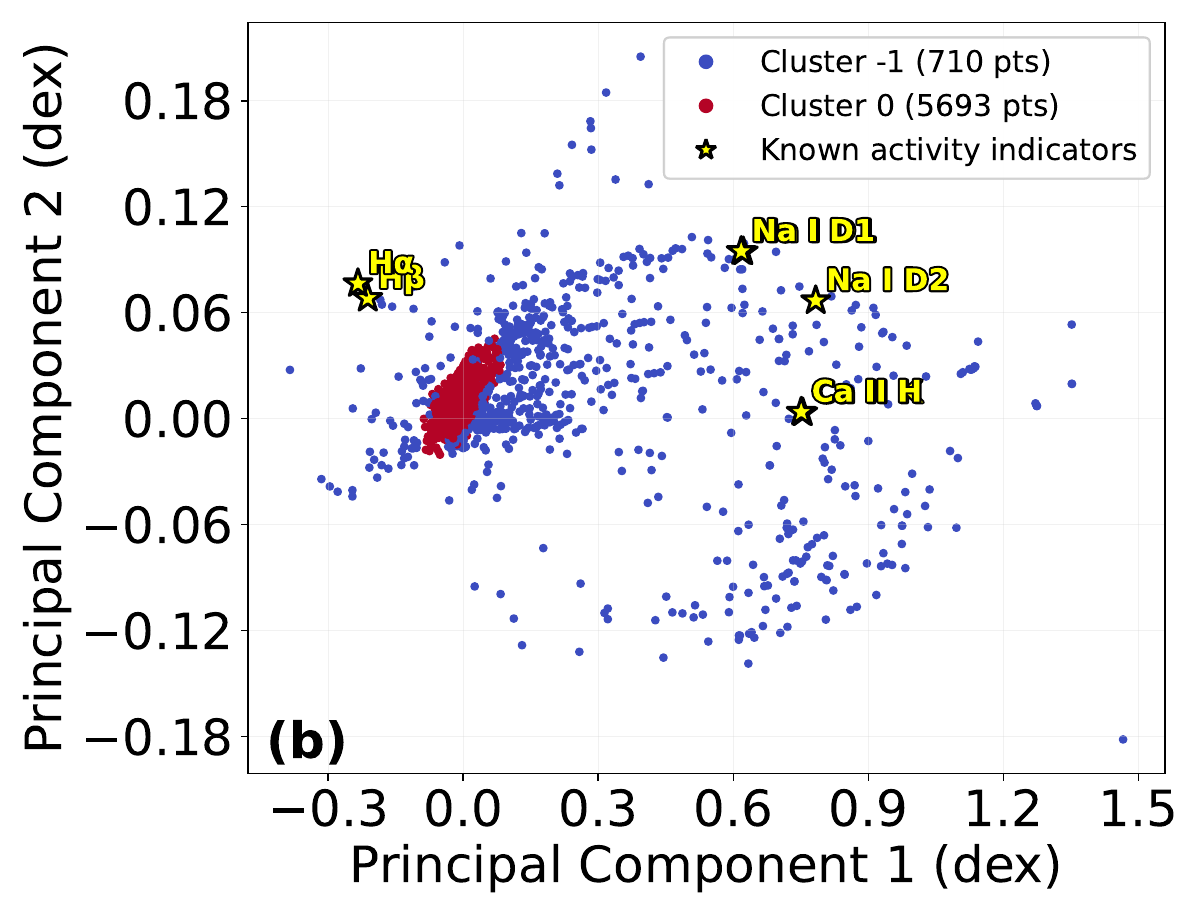}
\vspace{-0.5em}
\caption{PCA and DBSCAN results. \textbf{(a)} Weights for the first two PCA components versus chromospheric plateau temperature. \textbf{(b)} PC1--PC2 projection: stable core (red), activity-sensitive lines (blue), and known chromospheric indicators (yellow stars).}
\label{fig:pca_dbscan}
\end{figure}

\textbf{Dimensionality reduction.} PCA re-expresses each trajectory along orthogonal directions ordered by the variance they explain. We apply it to the $6403\times9$ LRD matrix. The first two principal components (``PC1'' and ``PC2'', respectively) provide a compact representation of the dominant forms of variability: PC1 explains 97.89\% of the variance and PC2 an additional 1.48\%. The clustering itself is performed in the full nine-dimensional LRD space. Figure~\ref{fig:pca_dbscan}(a) shows PC1 and PC2 weights as a function of chromospheric plateau temperature.

PC1 orders the lines by the signed size of their response: its weights are all positive and grow with chromospheric plateau temperature, so lines that weaken with activity project to positive PC1 and the few that deepen project to negative values. A small $|\mathrm{PC1}|$ therefore does not by itself imply stability. 
The weights of PC2 change sign along the sequence, negative up to the moderately active models and strongly positive for the hottest, so PC2 separates lines that respond early and then saturate from those that stay inert until the highest activity levels. 

\textbf{Clustering.} We explored multiple unsupervised approaches and focus here on DBSCAN \citep{Ester1996} because it identifies the dense stable core without imposing the number of clusters. Sensitivity is therefore defined by exclusion from that core: a line can be stable in one way, but can respond to chromospheric heating for several reasons.

Figure~\ref{fig:pca_dbscan}(b) shows DBSCAN clustering results in the PC1--PC2 projection. We use $\varepsilon=0.0129$ dex and $\mathrm{min\_samples}=10$, one of several combinations explored at this stage. A systematic criterion for fixing $\varepsilon$ is being developed for a forthcoming paper, so the exact population sizes should be read as indicative of the structure recovered rather than final.

At this $\varepsilon$, DBSCAN finds 5693 stable-core lines and 710 activity-sensitive lines, recovering known chromospheric indicators. In the PC1--PC2 projection, part of the sensitive population overlaps visually with the stable core and is separated only when the nine-dimensional trajectories are considered.

\section{Atomic properties and spectral sensitivity}

DBSCAN membership is strongly associated with lower-level energy. Activity-sensitive lines are concentrated at low $E_{\rm lower}$, with a median of 1.56~eV, whereas the stable core has a median of 3.69~eV. The overlap between the two populations shows that $E_{\rm lower}$ is a statistical discriminator, but not a unique predictor of activity sensitivity. Low-lying levels are heavily populated in the quiet model, and chromospheric heating depopulates them, so the opacity falls and the absorption weakens. The full NLTE response and the line-formation conditions must also be considered.

This trend agrees with \cite{Wise2022}, who found stronger activity-related variability in low-excitation Fe~I lines. They interpreted it, however, with an analytic photospheric description, in which the line response follows a Boltzmann excitation factor at a single formation temperature, whereas our classification uses NLTE models that explicitly include a chromosphere. We are investigating why lines with similar $E_{\rm lower}$ show different responses, through the formation region of the line core and the number of such regions. Ionization potential does not separate the groups: in Fe~I alone, 25\% of the lines are sensitive, and they sit at systematically lower $E_{\rm lower}$ than the rest of the species.

\begin{figure}
\centering
\includegraphics[width=.9\textwidth]{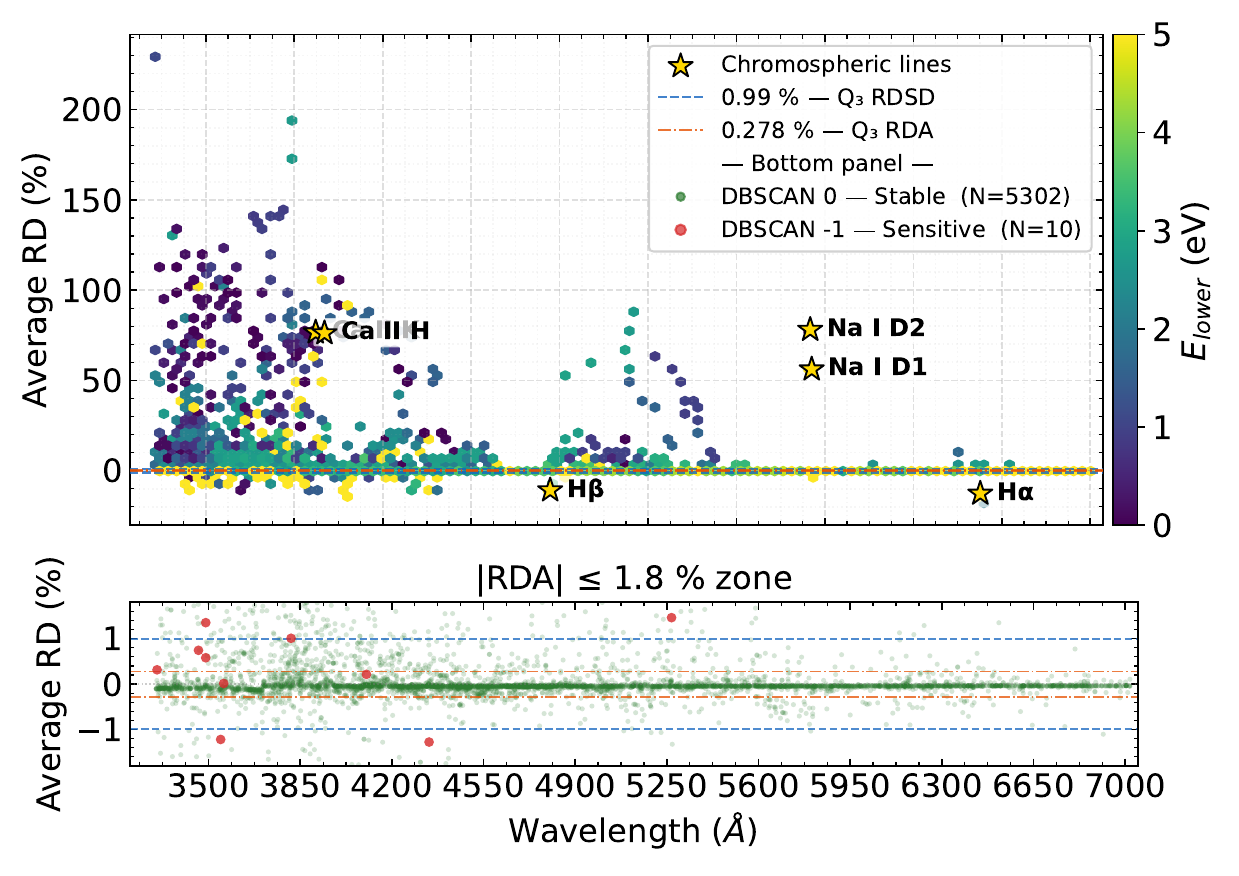}
\vspace{-1em}
\caption{Spectral sensitivity map. Upper panel: average RD (RDA) versus wavelength, colour-coded by $E_{\rm lower}$, with known chromospheric lines marked for reference. Lower panel: $|\mathrm{RDA}|\leq1.8\%$ zoom with DBSCAN membership.}
\label{fig:map}
\end{figure}

The spectral sensitivity map in Figure~\ref{fig:map} connects the model-by-model RD analysis of VZB25 with the present trajectory-based classification. It is built on the quantities an observer can measure directly: for each transition, RDA is the mean of the nine RD values and RDSD their standard deviation. Their third quartiles, $Q_3(\mathrm{RDA})=0.278\%$ and $Q_3(\mathrm{RDSD})=0.990\%$, mark the reference levels, lines that exceed them are concentrated toward the blue and near-UV.

The lower panel of the map zooms into the low-$|\mathrm{RDA}|$ range, where 5302 lines are classified as stable and 10 as activity-sensitive. These sensitive lines have a small average response, but their nine-model trajectories reveal threshold-like, nonlinear, or compensating behaviour. A small RDA therefore does not imply stability: averaging across the model sequence can hide variations that are recovered only when the full response pattern is retained.

\FloatBarrier

\section{Conclusions and Future Work}

Clustering the complete activity-response trajectory, not a single averaged value, identifies not only which lines are sensitive to chromospheric heating, but how and when that sensitivity emerges along the sequence. Linking these patterns to atomic properties yields two complementary products: a large set of activity-insensitive transitions for measurements that must be free of activity contamination, from radial velocities and abundances to stellar parameters, and a smaller set of activity-sensitive transitions that extend the chromospheric diagnostics beyond the canonical indicators.

One limitation is that the classification presented here is preliminary: the clustering criterion is still being consolidated, so the populations may shift, although the structure they recover is stable. A second is line blending, which can make a transition inherit the response of a close neighbour. NLTE diagnostics and integrated contribution functions are being used to identify these cases. The full catalogue will be published with the forthcoming paper.

The responses come from synthetic spectra computed at a step of $10^{-5}$~\AA\ with no instrumental or rotational broadening, as in VZB25, so they are upper bounds on what a real spectrograph would record. Matching them to the resolving power and coverage of current and forthcoming instruments is the next step, and it bears directly on the radial-velocity follow-up of PLATO Earth-like planet candidates. The longer-term goal is a line-by-line map of activity sensitivity that any high-precision spectroscopic programme can use, whether to avoid the affected transitions or to exploit them as activity diagnostics.

\end{document}